\documentclass[twocolumn,superscriptaddress,prl,showpacs]{revtex4}
\usepackage{graphicx,bm}
\usepackage[psamsfonts]{amssymb}
\usepackage{type1cm}
\usepackage{dcolumn}
\usepackage{amsmath}

\begin{document}

\title{Evidence for Dirac Fermions in a honeycomb lattice based on silicon}
\author{Lan Chen}
\affiliation{Institute of
Physics, Chinese Academy of Sciences, Beijing 100190, China}
\author{Cheng-Cheng Liu}
\affiliation{Institute of
Physics, Chinese Academy of Sciences, Beijing 100190, China}
\author{Baojie Feng}
\affiliation{Institute of
Physics, Chinese Academy of Sciences, Beijing 100190, China}
\author{Xiaoyue He}
\affiliation{Institute of
Physics, Chinese Academy of Sciences, Beijing 100190, China}
\author{Peng Cheng}
\affiliation{Institute of
Physics, Chinese Academy of Sciences, Beijing 100190, China}
\author{Zijing Ding}
\affiliation{Institute of
Physics, Chinese Academy of Sciences, Beijing 100190, China}
\author{Sheng Meng }
\affiliation{Institute of
Physics, Chinese Academy of Sciences, Beijing 100190, China}
\author{Yugui Yao\,\footnote{Corresponding author of the theoretical part, Email: ygyao@bit.edu.cn}}
\affiliation{School of Physics, Beijing Institute of
Technology,Beijing 100081, China}
\affiliation{Institute of
Physics, Chinese Academy of Sciences, Beijing 100190, China}
\author{Kehui Wu\,\footnote{Corresponding author, Email: khwu@aphy.iphy.ac.cn}}
\affiliation{Institute of Physics, Chinese Academy of Sciences,
Beijing 100190, China}
\date{\today}

\begin{abstract}
Silicene, a sheet of silicon atoms in a honeycomb lattice, was
proposed to be a new Dirac-type electron system similar as
graphene. We performed scanning tunneling microscopy and
spectroscopy studies on the atomic and electronic properties of
silicene on Ag(111). An unexpected $\sqrt{3}\times \sqrt{3}$
reconstruction was found, which is explained by an extra-buckling
model. Pronounced quasi-particle interferences (QPI) patterns ,
originating from both the intervalley and intravalley scattering,
were observed. From the QPI patterns we derived a linear
energy-momentum dispersion and a large Fermi velocity, which prove
the existence of Dirac Fermions in silicene.
\end{abstract}
\pacs{68.37.Ef, 73.22.-f, 61.48.-c, 71.20.Mq}

\maketitle

Group IV (Si, Ge) analogs of graphite have been discussed for a
long history even before the synthesis of isolated graphene
\cite{Takeda}, and recently there have been renewed interest in
this topic due to the novel concepts and applications brought by
graphene. The silicon version of graphene in which Si atoms
replace C atoms in a two-dimensional honeycomb lattice is named
silicene \cite{Aufray,Padova, Lalmi, Liu1, Cahangirov}.
Theoretical calculations show that silicene has also graphene-like
electronic band structure, supporting charge carriers behaving as
massless Dirac Fermions \cite{Liu1,Cahangirov}. Compared with
graphene \cite{Geim,Neto}, silicene has a larger spin-orbit
coupling strength, which may lead to larger energy gap at the
Dirac point and favor detectable quantum spin Hall effect
(QSHE)\cite{Liu1,Kane,Liu2}. Currently QSHE has been realized only
in HgTe-CdTe quantum wells and further studies have been hindered
by the challenging material preparation \cite{Bernevig}. The easy
preparation and compatibility with silicon-based nanotechnology
makes silicene particularly interesting for applications like QSHE
devices.

Despite the rapidly increasing amount of theoretical works on
silicene, there have been only a few experiments on silicene or
silicene nanoribbon \cite{Aufray,Padova,Lalmi}. Monolayer silicene
film has been successfully grown on Ag(111) \cite{Lalmi} and
ZrB$_2$ substrates \cite{Takamura}, and scanning tunneling
microscopy (STM) study revealed hexagonal honeycomb structure,
which was distinct from known surface structures of bulk silicon,
and resembles that of graphene \cite{Lalmi}. However, the reported
Si-Si distance of 17\% shorter than that for the bulk and for the
theoretical model implies unrealistic high compression of the
silicene lattice, which remains to be confirmed and understood
\cite{Lalmi}. Apart from the preparation and structural studies,
there is still no experiments on the electronic structure of
monolayer silicene. Experimentally establishing a common
understanding of the basic atomic and electronic properties of
silicene is therefore highly desirable.

In this Letter we report a study on silicene by low temperature
STM and scanning tunneling spectroscopy (STS). We observed an
unexpected $\sqrt{3}\times \sqrt{3}$ reconstruction on silicene
surface, in contrast to the 1$\times$1 structure reported
previously \cite{Lalmi}. Despite the structural difference, the
electronic property measured by STS is consistent with theory very
well. For examples, quasi-particle interference (QPI) patterns
suggesting intervalley and intravalley scattering of charge
carriers were observed, and a linear energy-momentum dispersion
relation and a large Fermi velocity were derived. Our results
provide a solid basis for further studies on electronic property
and device applications of silicene.

Experiments were performed in a home-built low-temperature STM
equipped with in-situ sample preparation facilities. A clean
Ag(111) surface was prepared by cycles of argon ion sputtering and
annealing. Silicon was evaporated from a heated Si wafer with a
deposition flux of about 0.05 ML/min, and the Ag(111) substrate
maintained at $\approx $500 K. The differential conductance
(dI/dV) spectra were measured as the in-plane ac component in the
tunneling current with a lock-in amplifier by superimposing an ac
voltage of 10 mV and 676 Hz on the given dc bias of the
substrate-tip gap. We performed STM measurements at both room
temperature and 77 K, and found no difference. All the data
presented in this paper were taken at 77 K.

\begin{figure}[tbp]
\includegraphics[width=8.5cm,angle=-0]{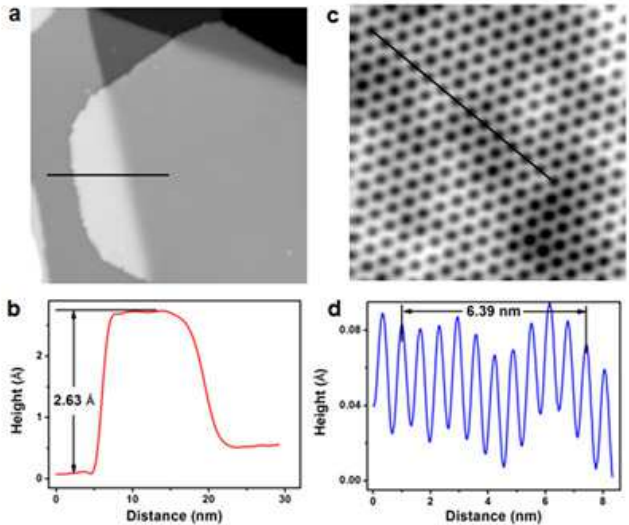}
\caption{ (color online) The STM image of a large area (65 nm 65
nm) consisting of a sheet of silicene on Ag(111) crossing two
substrate steps. (b) The line profile as indicated by the black
line in (a) shows that the island is of one atom thick. (c) The
high-resolution STM image (10 nm$\times$10 nm) of the silicene
surface taken at tip bias 1.0 V. The honeycomb structure is
clearly observed. (d) The line profile as indicated by the black
line in (c) showing both the lateral and vertical corrugation of
the structure observed by STM.}
\end{figure}

The growth of Si on Ag(111) is very sensitive to the Si coverage
and the substrate temperature \cite{Kara1}. We have explored a
wide range of Si coverage and substrate temperature, and several
metastable phases were found at low temperature range
\cite{Supplemental}. On the other hand, at sufficiently high
substrate temperature and small Si coverage, silicon routinely
forms one-atom thick, closely packed islands which we identify as
silicene sheets. Typical STM topographic image of the thus
obtained monolayer islands is shown in Fig. 1(a). The line profile
across the edge of a silicene island indicates that it has a
height of 2.63 \AA , corresponding to one atom thickness. Such
islands can run across steps of the Ag(111) substrate without
losing the continuity of the atomic lattice, similar to that of a
graphene sheet falling on a stepped surface. As expected, the high
resolution STM image in Fig. 1(c) shows a honeycomb structure
reflecting the three-fold symmetry of this film. However,
surprisingly, after a very careful calibration we derived a
periodic constant of 0.64$\pm $0.01 nm (shown in line profile in
Fig.1(d)), which is
approximately $\sqrt{3}$ a (a is the lattice constant of silicene, $\approx $%
0.38 nm, proposed in several theoretical works
\cite{Liu1,Cahangirov}). Such a large difference is impossible a
result of strain-induced lattice expansion of
compression. It is most likely that we observed a ($\sqrt{3}\times\sqrt{3}$%
)R30$^{\circ }$ superstructure, which frequently shows up in
three-fold symmetric systems.

To account for this observation, we have carried out
first-principles calculations using the projector augmented wave
(PAW) pseudopotential method and Perdew-Burke-Ernzerhof (PBE)
exchange-correlation potential \cite{Pedrew} implemented in the
VASP package \cite{Kresse}. The convergence criteria for energy
and force were set to 10$^{-5}$eV and 0.001eV/\AA , respectively.
We performed an extensive search for different geometric
configurations with ($\sqrt{3}\times \sqrt{3}$)R30$^{\circ }$
periodicity, and concluded that a ($\sqrt{3}\times
\sqrt{3}$)R30$^{\circ }$ superstructure cannot be stabilized in a
free-standing, fully relaxed silicene model. We also considered
the incorporation of Ag atoms in $\sqrt{3}\times\sqrt{3}$ sites in
the silicene lattice, but it was proven to be energetically
unfavorable. The period constant 0.64 nm cannot also be resulting
from the commensuration between the lattices of silicene and
Ag(111). Actually the theoretical modeling only gives a
$\sqrt{7}\times \sqrt{7}$ superstructure (with respect to the
1$\times $1 Ag substrate). We also exclude the possibility that
the honeycomb ($\sqrt{3}\times\sqrt{3}$)R30$^{\circ }$ structure
is simply an intervalley QPI pattern which was known in graphene
to create a ($\sqrt{3}\times\sqrt{3}$)R30$^{\circ }$ pattern
\cite{Niimi}. In fact, we observed such interference pattern (to
be shown in
the following context), but they can be distinguished from the ($\sqrt{3}%
\times \sqrt{3}$)R30$^{\circ }$ superstructure observed here, indicating
that they have different origins.

\begin{figure}[tbp]
\includegraphics[width=8.5cm,angle=-0]{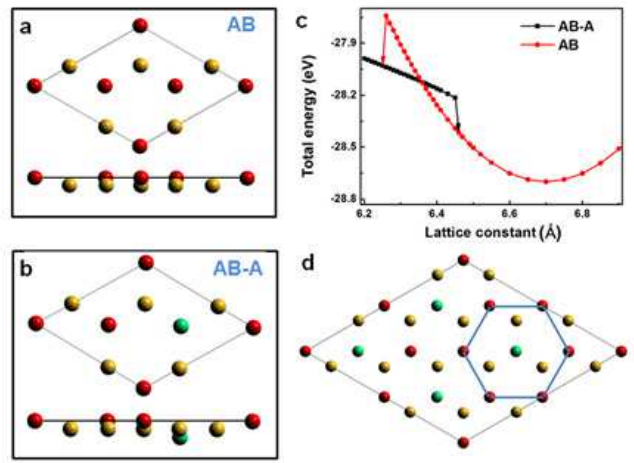}
\caption{ (color online) (a), (b) The top view and side view of
the lattice geometry of the $\sqrt{3}\times\sqrt{3}$
superstructure (AB\={A}) and low-buckled phase (AB), respectively.
Note that in the $\sqrt{3}\times\sqrt{3}$ superstructure a Si atom
with planar coordinate (2/3, 2/3) is pulled downward. The red,
yellow and green balls represent the A, B and \={A} Si atoms,
respectively. (c)The structural phase transition diagram of
silicene depending on the lattice constants of
$\sqrt{3}\times\sqrt{3}$ superstructure. (d) A larger schematic
model illuminating the honeycomb structure of
$\sqrt{3}\times\sqrt{3}$ reconstructed silicene.}
\end{figure}

Finally, we found a clue by comparing the obtained periodicity,
0.64 nm, with the theoretically proposed values, all around 0.67
nm \cite {Liu1,Cahangirov,Ding}. The experimental value is about
4\% smaller than the theoretical ones, indicating that the
substrate may exert some influence to result in the contraction of
the silicene lattice. Although the detail mechanism for such
contraction still remains to be understood, for a phenomenological
model we can fix the periodicity constant in a series of
contracted silicene, and fully relax the inner coordinates of Si
atoms. Interestingly, we found that the ($\sqrt{3}\times
\sqrt{3}$)R30$^{\circ }$ reconstruction can indeed be stabilized
in a contracted lattice. The Si atoms in previous theoretical
low-buckled silicene (Fig. 2(a)) have two different heights, which
we named AB configurations. To understand the experimentally
observed ($\sqrt{3}\times \sqrt{3}$) superstructure, we considered
a configuration with extra buckling of the Si atom with planar
coordinate (2/3, 2/3) downward in one ($\sqrt{3}\times \sqrt{3}$)
silicene unit cell, which is named AB\={A} configuration, as shown
in Fig. 2(b). We carried on a comprehensive study of the structure
and stability of a series of contracted silicene structures from
the AB and AB\={A} configuration with and without small
perturbation as the initial structures, respectively. Fig. 2(c)
shows the structure phase transition diagram of the
($\sqrt{3}\times \sqrt{3}$) superstructure and the low-buckled AB
phase with minimum energy and stability, through structural
optimization by keeping the periodicity constant in the contracted
region. When the lattice constant is smaller than 6.25 \AA , both
the AB\={A} configuration and the AB configuration as the initial
structures return to the ($\sqrt{3}\times \sqrt{3}$)
superstructure (AB\={A} configuration). In contrast, when the
lattice constant is larger than 6.45 \AA , an initial AB\={A}
configuration will spontaneously transfer to the low-buckled AB
configuration. The $\sqrt{3}\times \sqrt{3}$ superstructure
(AB\={A} configuration) is robust and stable when the lattice
constant is smaller than 6.35 \AA . Taking into account the scale
error, our experimental silicene sample may be in the AB\={A}
phase with the ($\sqrt{3}\times \sqrt{3}$) superstructure.

Silicene has been predicted to have Dirac-type electron structure
around the Fermi energy \cite{Liu1,Cahangirov}, similar to that of
graphene. This however remains to be experimentally confirmed. On
the other hand, measuring the characteristic electronic structure
of our silicene film can also provide an additional proof of the
basic atomic structure of our film. For graphene, the Dirac cones
are located at high symmetric K points in the 2D Brillouin zone.
The constant energy contours in reciprocal space cut through the
electron or hole conical sheets near $E_F$ resulting in small
circles centered at the K points (shown in Fig. 3(d)). Free
carriers can be scattered within the small circles (intravalley
scattering) or between circles (intervalley scattering)
\cite{Rutter,Brihuega,Xue}, resulting in QPI patterns in real
space, which have been observed STS. Similar QPI patterns are
expected to exist for silicene based on the same analysis.

\begin{figure}[tbp]
\includegraphics[width=8.5cm,angle=-0]{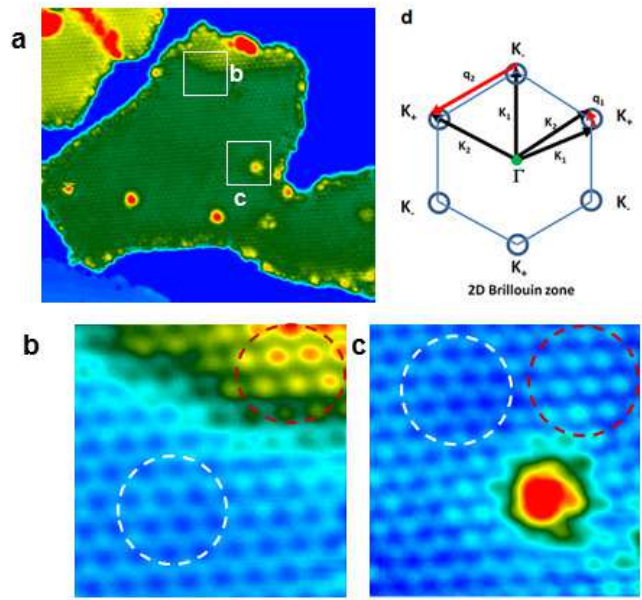}
\caption{ (color online) (a) The STM image (50 nm $\times$45 nm)
of a silicene island taken at tip bias -1.1 V. The white boxes
label two typical defect sites, namely step edge and point defect.
where scattering patterns with hexagonal close packing structures
were observed, as enlarged in (b) and (c). (d) Schematic of 2D
Brillouin Zone (Blue lines), constant energy contours (grey rings)
at K points, and the two types of scattering vectors: $q_1$
(intravalley, short red arrow) and $q_2$ (intervalley, long red
arrow). }
\end{figure}

We have observed short-wavelength interference patterns in STM
images of silicene, consistent with an intervalley scattering.
From Fig. 3(d) one can see that the wave vectors of intervalley
scattering, $q_2$ , is close in length to the wave factor $\kappa
_{1(2)}$, which has a length of 1/$\sqrt{3}$ in reciprocal space, corresponding to a ($\sqrt{3}\times \sqrt{3%
}$)R30$^{\circ }$ periodicity in real space. Such a
($\sqrt{3}\times\sqrt{3}$)R30$^{\circ }$ interference pattern had
been observed in areas near step edges and defects which serve as
scattering centers, as illustrated in Fig. 3(a)-(c). The red
circle in (b) is drawn near the step edge, and in (c) it is along
the close-packing direction away from the point defect. Within the
red circles the surface is imaged as close-packed protrusions. In
contrast, within the white circles the surface is imaged as
honeycomb structure, as commonly observed in large area silicene
islands. Therefore, we can find a clear phase shift between
the atomic corrugation and the QPI pattern, although both of them exhibit ($\sqrt{3}\times\sqrt{3}$)R30$%
^{\circ }$ periodicity. The QPI pattern originating from the
scattering center extends for a length of only a few nm. Such a
characteristic decaying length cannot make us assign the
($\sqrt{3}\times\sqrt{3}$)R30$^{\circ }$ structure observed on
larger islands simply as an QPI pattern. The observation of QPI
patterns, consistent with analysis based on the theoretical band
structure of silicene is an additional proof that the underlying
atomic structure of our film is graphene-like, with only some
buckling that does not change the basic electronic structure of
the film.

To investigate the electronic structure of silicene in more
detail, we performed STS measurements (dI/dV curves and maps) on
the film. Typical dI/dV curve taken at 77 K is shown in Fig.4(a).
Beside the pronounced peak at 0.9 V, there is a small dip located
at about 0.5 V, which is attributed to the position of Dirac point
(DP) of silicene. The dip in dI/dV curve corresponding to DP is
not much obvious compared with that of graphene \cite{Zhang,Zhao},
which is due to the pronounced electronic density of states (DOS)
of the underlying Ag(111) substrate superimposed in the dI/dV
spectra. The Fig. 4(c)-(e) are dI/dV maps of a silicene island
(second layer of silicene) on the single layer of silicene,
reflecting the distribution of LDOS in real space. Wave-like QPI
patterns near the boundary of the island were clearly observed.
The wavelength changes as a function of tip bias voltage. As the
bias increases from -0.4 V to -1.1 V, the wavelength decreases
correspondingly from 2.8 nm to 1.6 nm.

\begin{figure}[tbp]
\includegraphics[width=8.5cm,angle=-0]{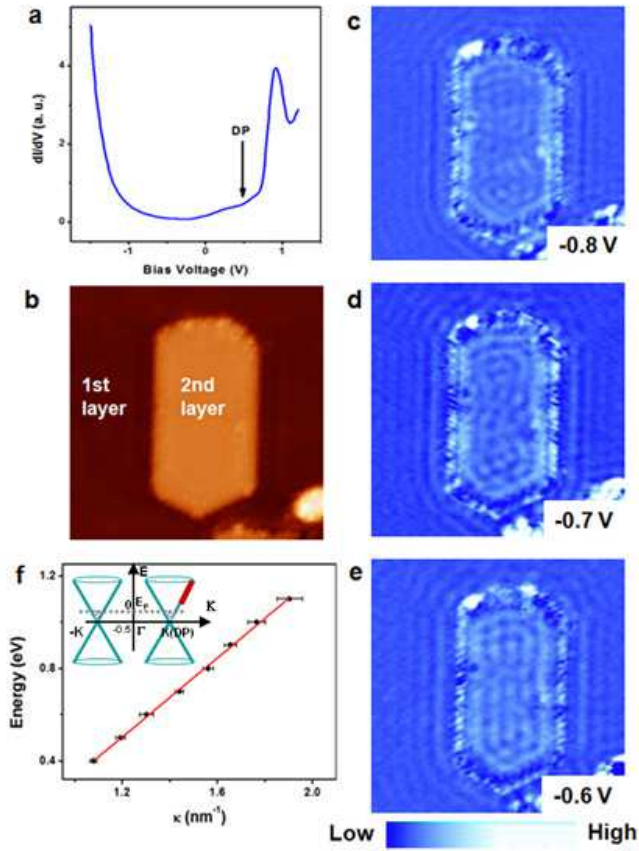}
\caption{ (color online) (a) dI/dV curves taken at 77 K. The
position of Dirac point (DP) is labeled. (b) The STM image (40 nm
$\times$ 40 nm) of 1ML silicene surface containing an island of
second layer taken at tip bias -1.0 V. (c), (d) and (e) dI/dV maps
of the same area as (b) taken at tip bias -0.8 V, -0.7 V and -0.6
V, respectively. (f) Energy dispersion as a function of $\kappa$
for silicene determined from wavelength of QPI patterns. The inset
shows an schematic drawing of the overall band structure, with the
relative location of DP, $E_F$ and our data points (red line)}
\end{figure}

The 2D constant-energy contours in reciprocal space (Fig. 3(d))
are used to understand the QPI patterns. The wave factor $q_1$ of
intravalley scattering connect points within a single
constant-energy circle and determine the observed long wavelength
interference pattern. In order to deduce the quasipartical
energy-momentum dispersion relation, we drew E($\kappa $) curve in
Fig. 4(f), where $\kappa $ is the radius of constant-energy circle
at K point with 2$\kappa $= $q_1$. The values of $q_1$ are
determined by measuring the wave length of QPI patterns in dI/dV
maps. We found $\kappa$ varied linearly with
energy, with Fermi velocity $V_F=(1.2\pm 0.1)\times 10^6$ m/s. The $\kappa $%
=0 energy intercept gives the Dirac energy, E$_F$-E$_D$ = 0.52$\pm
$0.02 eV, in consistent with the position of DP in dI/dV spectra
(Fig. 4(a)) very well. The linear E-$\kappa$ dispersion proves the
existence of Dirac cone in electronic band structures of silicene.
The surprising large Fermi velocity, comparable with that of
graphene \cite{Rutter,Berger}, suggests the prospective
applications comparable with those have been proposed or realized
in graphene.

In summary, we have obtained silicene film on Ag(111) and observed
an unexpected $\sqrt{3}\times\sqrt{3}$ buckled superstructure. QPI
patterns resulting from both intervalley and intravalley
scattering were found, and the study of the linear energy-momentum
dispersion and the measured Fermi velocity, as high as $10^6$ m/s,
proves that quasiparticles in silicene behave as massless Dirac
fermions. Based on these results, many further interesting studies
may be proceed, including gap opening in silicene, which can
realize the quantum spin Hall effect \cite{Liu1}, and high
temperature electron-phonon superconductivity in hydrogenated
silicene \cite{Kara2,Savini}.

\textit{Acknowledgements}: We thank Prof. Y.Q. Li and Prof. Min
Qiu for helpful comments. This work was supported by the NSF of
China (Grants No. 11174344, 10974231, 11174337, 11074289), and the
MOST of China (Grants No. 2012CB921700, 2011CBA00100).


\begin{thebibliography}{99}
\bibitem{Aufray}  B. Aufray, A. Kara, S. Vizzini, H. Oughaddou, C. L\.{e}%
andri, B. Ealet, G. L. Lay, Appl. Phys. Lett. \textbf{96}, 183102(2010).

\bibitem{Verri} G. G. Guzm\'an-Verri and L. C. L. Voon, Phys. Rev. B \textbf{76},
075131 (2007)

\bibitem{Padova}  P. D. Padova \textit{et al.}, Appl. Phys. Lett. \textbf{96}, 261905(2010).

\bibitem{Lalmi}  B. Lalmi, H. Oughaddou, H. Enriquez, A. Kara, S. Vizzini,
B. Ealet, B. Aufray, Appl. Phys. Lett. \textbf{97}, 223109(2010).

\bibitem{Liu1}  C. C. Liu, W. X. Feng, Y. G. Yao, Phys. Rev. Lett. \textbf{%
107}, 076802(2011).

\bibitem{Cahangirov}  S. Cahangirov, M. Topsakal, E. Akturk, H. Sahin, S.
Ciraci, Phys. Rev. Lett. \textbf{102}, 236804(2009).

\bibitem{Takeda} K. Takeda and K. Shiraishi, Phys. Rev. B
\textbf{50}, 14916(1994).

\bibitem{Geim}  A. K. Geim, K. S. Novoselov, Nature Mater. \textbf{6},
183(2007).

\bibitem{Neto}  A. H. Castro Neto, F. Guinea, N. M. R. Peres, K. S.
Novoselov, A. K. Geim, Rev. Mod. Phys. \textbf{81}, 109(2009).

\bibitem{Kane}  C. L. Kane, E. J. Mele, Phys. Rev. Lett. \textbf{95},
226801(2005).

\bibitem{Liu2}  C. C. Liu, H. Jiang, Y. G. Yao, Phys. Rev. B \textbf{84},
195430(2011).

\bibitem{Bernevig} B. A. Bernevig, T. L. Hughes, and S. C. Zhang, Science
\textbf{314}, 1757(2006).

\bibitem{Takamura} Y.Takamura, private communication.

\bibitem{Kara1}  A. Kara, H. Enriquez, A. P. Seitsonen, L. C. Lew Yan Voon,
S. Vizzini, B. Aufray, H. Oughaddou, Surf. Sci. Rep. \textbf{67}, 1(2012).

\bibitem{Supplemental}  See Supplemental Materials at http://.

\bibitem{Pedrew}  J. P. Perdew, K. Burke, M. Ernzerhof, Phys. Rev. Lett.
\textbf{77}, 3865(1996).

\bibitem{Kresse}  G. Kresse J. Furthm\"{u}ller, Phys. Rev. B \textbf{54}, 11169(1996).

\bibitem{Niimi}  Y. Niimi, T. Matshui, H. Kambara, K. Tagami, M. Tsukada, H.
Fukuyama, Phys. Rev. B \textbf{73}, 085421(2006).

\bibitem{Ding}  Y. Ding, J. Ni, Appl. Phys. Letts. \textbf{95}, 083115(2009).

\bibitem{Rutter}  G. M. Rutter, J. N. Crain, N. P. Guisinger, T. Li, P. N.
First, J. A. Stroscio, Science \textbf{317}, 219(2007).

\bibitem{Brihuega}  I. Brihuega \textit{et al.}, Phys. Rev. Letts.
\textbf{101}, 206802(2008).

\bibitem{Xue}  J. Xue, J. Sanchez-Yamagishi, K. Watanabe, T, Taniguchi, P.
Jarillo-Herrero, B. J. LeRoy, Phys. Rev. Letts. \textbf{108}, 016801(2012).

\bibitem{Zhang}  Y. B. Zhang, V. W. Brar, F. Wang, C. Girit, Y. Yayon, M.
Panlasigui, A. Zettl, M. F. Crommie, Nature Phys. \textbf{4}, 627(2008).

\bibitem{Zhao}  L. Zhao et al., Science \textbf{333}, 999(2011).

\bibitem{Berger}  C. Berger et al., Science \textbf{312}, 1191(2006).

\bibitem{Kara2}  A. Kara, C. Leandri, M. E. Davila, P. De Padova, B. Ealet,
H. Oughaddou, B. Aufray, G. Le Lay, J. Supercond. Nov. Magn. \textbf{22},
259(2009).

\bibitem{Savini}  G. Savini, A. C. Ferrari, F. Giustino, Phys. Rev. Letts.
\textbf{105}, 037002(2010).
\end{thebibliography}
\end{document}